\documentstyle[aps,prd,epsfig]{revtex}
\newcommand{\s}{$\sigma_{tot}^{pp}$ }

\tightenlines

\begin{document}

\title{\bf High-Energy Proton-Proton Forward Scattering and Derivative
Analyticity Relations}

\author{R. F. \'Avila$^{1}$, E. G. S. Luna$^{2}$ and M. J. Menon$^{2}$\\
\small\it $^1$Instituto de Matem\'atica, Estat\'{\i}stica e Computa\c c\~ao
Cient\'{\i}fica\\
Universidade Estadual de Campinas, UNICAMP\\
13083-970 Campinas, SP, Brazil \\
\small\it $^2$Instituto de F\'{\i}sica Gleb Wataghin\\ 
Universidade Estadual de Campinas, UNICAMP\\
13083-970 Campinas, SP, Brazil}

\maketitle

\begin{abstract}

We present the results of several parametrizations to two different
ensemble of data on $pp$ total cross sections (\s) at the highest
center-of-mass energies (including cosmic-ray information). The results are 
statistically consistent with two distinct scenarios at high energies. From
one ensemble the prediction for the LHC ($\sqrt s = 14$ TeV) is 
$\sigma_{tot}^{pp} = 113 \pm 5$ mb and from the other, $\sigma_{tot}^{pp} =
140 \pm 7$ mb.  From each parametrization, and making use of derivative
analyticity relations (DAR), we determine $\rho(s)$ (ratio between the forward
real and imaginary parts of the elastic scattering amplitude). A discussion on 
the optimization of the DAR in terms of a free parameter is also presented.
In all cases good descriptions of the experimental data are obtained.

\end{abstract}

\section{Introduction}

 The total cross section, $\sigma_{tot}$, and the $\rho$ parameter
(ratio of the real to imaginary part of the forward elastic
scattering amplitude) are important quantities in the investigation
of elastic hadron-hadron scattering at high energies \cite{matthiae,blockcahn}.
They are expressed in terms of the elastic scattering amplitude,
$F(s,t)$, by the formulas,

\begin{equation}
\sigma_{tot}(s) = 4\pi Im F(s,t=0), 
\end{equation}

\begin{equation}
\rho(s) = \frac{Re F(s,t=0)}{Im F(s,t=0)},
\end{equation}
where $\sqrt s$ is the center-of-mass energy and $t$ the four-momentum transfer
squared.

For {\it proton-proton ($pp$) collisions}, both quantities have been extracted
from accelerator experiments and the results extend up to $\sqrt s \sim 63$
GeV. On the other hand, \s may also be inferred from cosmic-ray  experiments
at still higher energies, $\sqrt s \sim 10$ TeV. However, these cosmic-ray
results are model-dependent, since they are obtained from proton-air cross
sections \cite{engel} and this has originated different results for \s in the
interval $\sqrt s = 5 - 25$ TeV, which exhibit a reasonable degree of
discrepancy. 

Several models present extrapolations at the cosmic-ray region and the observed
discrepancies may be accommodated by different models \cite{mm,dl,dgp,bsw,block}. 
 As we shall recall, presently, it is difficult to decide which could
be the ``correct'' result and it seems that the general trend is to expect 
the new data from the next accelerator experiments, the BNL Relativistic Heavy
Ion Collider (RHIC), $\sqrt s \sim 500$ GeV \cite{rhic}  and the CERN
Large Hadron Collider (LHC), $\sqrt s \sim 14$ TeV \cite{totem}.

However, at this stage, we understand that a model independent
analysis of the experimental information presently available, taking detailed
account of the discrepancies and its consequences, may bring important insights
on the subject. This is the central point we are interested in.

In this communication, we first investigate two sets of distinct results
for \s, at cosmic-ray energies, in a model independent way that also takes
into account of the 
experimental data at the accelerator region ($\sqrt s > 10$ GeV). In each
case we fit four different analytic functions to each ensemble of data,
using the CERN-MINUIT routine \cite{minuit}.
Next we make use of the one-subtracted derivative
analyticity relation (DAR) in order to obtain analytic parametrizations for the
$\rho$ parameter as function of the energy, from both ensembles and for all the
parametrizations. In this investigation we stress that the {\it general}
expression of the DAR has a free parameter and we present a study on the
practical applicability of this parameter in the fits to the $\rho$ data. In
all cases good descriptions of the experimental data are obtained.

The paper is organized as follows. In Sec. II
we review the experimental information on \s from accelerator and
cosmic-ray experiments and recall the main steps connecting integral and
derivative  analyticity relations. The fits concerning
\s and the results for $\rho(s)$ are presented in Sec. III. In Sec. IV we
discuss in some detail all the results obtained and present comparisons
with some model predictions. The conclusions and some final remarks are the
content of Sec. V.

\section{Experimental Information and Analytical Approach}

Here we first review the experimental information presently available on
\s at the highest energies, stressing the puzzles involved at the cosmic-ray
region. We also recall some essential formulas connecting integral and
derivative analyticity relations and the corresponding derivative relation
between \s and $\rho$ (the analytical approach).

\subsection{Experimental information on $pp$ total cross sections}

As mentioned before, in order to analyze the experimental information
presently available on $\sigma_{tot}(s)$ for $pp$ interaction
at energies
beyond $\sim 10$ GeV (high energy region), we must distinguish between
accelerator and cosmic-ray information.  In the later case we follow a
discussion first presented in Ref. \cite{mm}.

Concerning accelerators, data on \s from three experiments at the
CERN Intersecting Storage Ring (ISR),
between $23.5$ GeV and $62.3$ GeV, were critically analyzed by Amaldi
and Schubert and we shall consider here the final mean values from their
analysis \cite{pp}. We also include the results at $13.8$ GeV
and $19.4$ GeV, obtained in Fermilab \cite{pp}. 

Although other accelerator data exist in this region, this set is suitable for
the analyzes we are interested in. From one hand, these data represent the
correct trend of $pp$ collisions in the region $10 - 60$ GeV and allow a
statistical approach that includes different cosmic-ray information (see what
follows). On the other hand, as will be discussed in Sec. IV, they are
adequate for comparison with a model that predicts a faster rising of \s than
generally expected.

Differently from accelerator data, cosmic-ray experimental information
on \s comes from proton-air cross sections. We recall that
antiprotons are not expected to  have any significant role in these
interactions and therefore the bulk of cosmic ray information
on hadron-nucleus cross sections does not concern antiprotons.

 Now, since what is extracted in these experiments is the proton-air cross section, the 
determination of the $pp$ cross section depends on nuclear model 
assumptions \cite{engel,mm} and this has originated some puzzles, as reviewed
in what follows.

The first information on \s from cosmic ray experiments, at $\sqrt s = 30$
TeV, was reported by Baltrusaitis {\it et al.} in 1984 \cite{fly}. Extensive
air showers recorded by Fly's Eye detector allowed the determination of the
proton-air inelastic cross section. Based on Glauber theory, assuming a
Gaussian profile for the nucleus and the proportionality between total cross
section and the slope parameter (geometrical scaling), the authors inferred
\cite{fly}

\begin{equation}
\sigma_{tot} = 120 \pm 15 \ \textnormal{mb} \quad {\rm at} \quad \sqrt s = 30 \
\textnormal{TeV}.
\end{equation}

In 1987, based also on data from Fly's Eye experiment, Gaisser, Sukhatme and
Yodh  (GSY) introduced the limit $\sigma_{tot}^{pp} \geq  130 \ \textnormal{mb}
\quad {\rm at} \quad \sqrt s \sim 30$ TeV and, taking account of various
process in the Glauber theory and additional assumptions,  estimated \cite{gsy}

\begin{equation}
\sigma_{tot}^{pp} = 175^{+40}_{-27}\ \textnormal{mb} \quad {\rm at} \quad \sqrt
s = 40\ \textnormal{TeV}.
\end{equation}

Afterwards, extensive air shower data collected in the Akeno observatory allowed new
estimates in the range $5-25$ TeV, reported by Honda {\it et al.}
\cite{akeno}, which is in agreement with the result reported by Baltrusaitis
{\it et al.} and in  disagreement with the GSY result. In particular,
extrapolations by Honda {\it et al.} indicated \cite{akeno}

\begin{equation}
\sigma_{tot}^{pp} = 133 \pm 10 \ \textnormal{mb} \quad {\rm at} \quad \sqrt s =
40 \ \textnormal{TeV},
\end{equation}
which is in agreement with the result reported by Baltrusaitis {et al.} and in 
disagreement with the GSY result. In the same year, Nikolaev claimed that the
Akeno results have been underestimated by about $30$ mb \cite{niko} and, if we
take this correction into account, the data in the interval $5-25$ TeV show
agreement with the early GSY result.  An important point is the fact that the
analysis by Nikolaev seems correct, has never been criticized and the same is
true for the GSY limit and result. 

All these cosmic-ray 
informations are shown in Figure 1, together with the accelerator data
at lower energies. We stress that these experimental information concern only
$pp$ and not $\overline{p}p$ interactions. From this Figure we clearly see the
discrepancies between Honda/Baltrusaitis and GSY/Nikolaev.

For these reasons, in this paper we  investigate the behavior of the
total cross section taking account of the discrepancies that
characterize the cosmic ray information.
In order to do that, we consider {\it two ensembles of data and experimental
information}, with the following notation:

\vspace{0.3cm}

Ensemble A: accelerator data + Honda {\it et al.} + Baltrusaitis
{\it et al.}

\vspace{0.3cm}

Ensemble B: accelerator data + Nikolaev + GSY

\vspace{0.3cm}

For ensemble A the references are \cite{pp,fly,akeno} and for
ensemble B, \cite{pp,gsy,niko}. In Sec. III we make use of
different parametrizations to fit the data from  each ensemble.

\begin{figure}
\begin{center}

\epsfig{figure=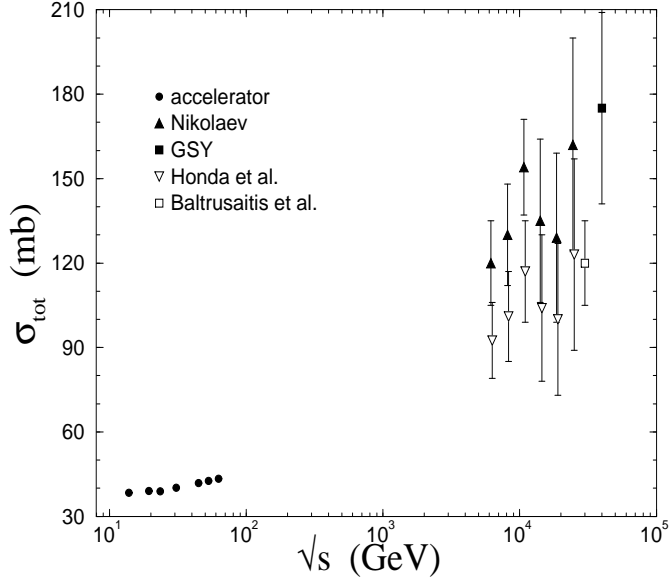,width=10cm,height=9cm}
\vspace{0.5cm}

\caption{Experimental information on $pp$ total cross sections:
accelerator data in the interval $13.8 < \sqrt s < 62.5$ GeV
and cosmic-ray results in the interval $6.3\ \textnormal{TeV} < \sqrt s <
40$ TeV.}
\end{center}
\end{figure}

\vspace{0.3cm}

\subsection{Derivative analyticity relations and the $\rho $ parameter}
\vspace{0.3cm}

Both $\sigma_{tot}(s)$ and $\rho(s)$ play a central role in the investigation
of high-energy hadron scattering. Due to the connections between forward
real and imaginary parts of the scattering amplitude, Eqs. (1) and (2),
the analyticity (dispersion) relations constitute a suitable, model
independent, approach for a simultaneous study of these quantities.

 Although integral dispersion relations have been widely used in the study
of hadronic scattering, in general, the analytical and/or numerical
integrations are not an easy task. However, at sufficiently high energies, the
smooth increase of \s (Fig. 1) allows to connect the integral form with a
derivative one, which is easier to handle.
The so called derivative analyticity relations (DAR) were introduced in the
seventies \cite{dar} and since then have been critically investigated
\cite{crit-dar,kolarfischer}. Recently, a recursive approach was
developed, as well as generalizations to an arbitrary number of subtractions,
for both cross even and odd amplitudes, near the forward direction \cite{mmp}.
 As in the integral case, the convergences may be controlled by subtractions
and specific formulas are associated with cross even and odd functions
(scattering amplitudes in the case of particle-particle and anti
particle-particle interactions) \cite{blockcahn}.

At this point we stress once more that we are only interested in $pp$
scattering (where discrepancies happen in $\sigma_{tot}(s)$) and not
$\overline{p}p$. Besides, our investigation concerns the highest energies,
characterized by the smooth increase of \s, nearly as a power on $ln s$. For
these reasons in our analyzes we will consider only an even amplitude, as the
leading contribution, and only one subtraction.

A detailed
deduction on how to obtain DAR from integral relations may be found in
references \cite{dar,crit-dar,kolarfischer,mmp}. Here we only review the main
steps concerning our case of interest, namely, one subtraction and an even
amplitude. We begin with the well known once-subtracted integral dispersion
relation (even amplitude) in the forward direction ($t=0$)
\cite{blockcahn,byronfuller}

\begin{equation}
Re f_{+}(s)=  K + \frac{2s^{2}}{\pi}I,
\end{equation}
where $K \equiv Re f_{+}(0)$ is the subtraction constant and
\begin{equation}
I=P\!\int_{s_{0}}^{+\infty} ds' 
\frac{1}{s'(s'^{2}-s^{2})}Imf_{+}(s').
\label{intI}
\end{equation}

Following Bronzan, Kane and Sukhatme \cite{dar}, and also \cite{mmp}, we
consider a real  parameter $\alpha$ so that after multiply and divide by
$s^{\alpha}$ and integrating by parts we obtain

\begin{eqnarray}
I&=&\frac{1}{2ss'}\ln\left| \frac{s'-s}{s'+s}\right| Im f_{+}(s')
{\large \mid}_{s'=s_{0}}^{\infty} - \nonumber
\\
&-&\frac{1}{2s}\int_{s_{0}}^{\infty}ds' s'^{\alpha -1}
\ln\left| \frac{s'-s}{s'+s}\right| \left(\frac{\alpha -1}{s'} + \frac{d}{ds'}\right)
Im f_{+}(s')/s'^{\alpha}.
\label{aprox}
\end{eqnarray}
Taking account of the high-energy region $(s\gg s_{0} \sim m^2 \sim 1$
GeV$^2$) and performing  a change of variable $ s' = e^{\xi '}, \:\:\: s =
e^{\xi}, $
the last equation may be put in the form

\begin{equation}
I=\frac{1}{2s}\int_{\ln s_{0}}^{\infty}d\xi ' s'^{\alpha -1}
\ln \coth\frac{1}{2}|\xi - \xi '|\left(\alpha -1 + \frac{d}{d\xi '}\right)
Im f_{+}(s')/s'^{\alpha}.
\label{25}
\end{equation}

Expanding $Im f_{+}(s')/s'^{\alpha}$ in powers of 
$\xi ' -\xi$, after some manipulation and taking account of the
high energy limit ($s_{0}\rightarrow 0$, that is, $\ln s_{0}
\rightarrow -\infty$) we obtain

\begin{equation}
Ref_{+}(s)= K + s^{\alpha}\sum_{n=0}^{\infty}\frac{d^{(n)}}{d \ln s^{(n)}}
\left( Im f_{+}(s)/s^{\alpha}\right)
\frac{I_{n}}{n!}\:\: ,
\label{29}
\end{equation}
where $I_{n}$ represents the integral in the variable $\xi '$,

\begin{equation}
I_{n}=\frac{1}{\pi}\int_{-\infty}^{+\infty}d\xi 'e^{(\alpha -1)(\xi '-\xi)}
\ln\coth \frac{1}{2}|\xi ' -\xi|\left(\alpha -1+\frac{d}{d\xi '}\right)
(\xi '-\xi)^{n}.
\nonumber
\end{equation}

We have assumed that  the series may be integrated term by term.
Denoting $\xi ' - \xi\equiv y$ and integrating by parts, this equation may be
put in the form

\begin{equation}
I_{n}=\frac{1}{\pi} \ln\coth\frac{1}{2}|y|\:\: e^{(\alpha -1)y}y^{n}
{\large \mid}_{-\infty}^{+\infty}
+ \frac{1}{\pi}\int_{-\infty}^{+\infty}dy \frac{e^{(\alpha -1)y}}{\sinh y}y^{n}.
\label{32}
\end{equation}

A central point here is that the first term on the right hand side
{\it converges to zero only for}
\begin{equation}
0 < \alpha < 2.
\end{equation}
In this case we only have the second term, which can be expressed as
a recursive relation in terms of the parameter $\alpha$ \cite{mmp}:
\begin{equation}
I_{n}=\frac{d^{(n)}I_{0}}{d\alpha^{(n)}},
\end{equation}
and integration in the complex plane gives
\begin{equation}
I_{0}= \tan \left(\frac{\pi}{2}(\alpha -1)\right).
\end{equation}

With this, we obtain an expression connecting the real part of an even amplitude with
the derivatives of the imaginary part at the same energy, namely the DAR:

\begin{equation}
\frac{Ref_{+}(s)}{s^{\alpha}} = \frac{K}{s^{\alpha}} + \tan 
\left[ \frac{\pi}{2}\left(\alpha -1 +\frac{d}{d\ln s}\right) \right]
\frac{Imf_{+}(s)}{s^{\alpha}} .
\label{40}
\end{equation}

The leading term in the tangent series reads
 
\begin{equation}
\tan \left( \frac{\pi }{2}(\alpha
-1)\right) \frac{Im f_{+}(s)}{s^{\alpha }}+\frac{\pi }{2}\sec
^{2}\left( \frac{\pi }{2}(\alpha -1)\right) \frac{d}{d\ln s}\left( \frac{
Im f_{+}(s)}{s^{\alpha }}\right),
\end{equation}
and making use of the normalization ($k^2 \sim s$),

\begin{equation}
\frac{f_{+}(s)}{s} \equiv F(s,t=0),
\end{equation}
we obtain the general result for a forward amplitude:

\begin{eqnarray}
Re F(s,0)= \frac{K}{s} + \tan \left( \frac{\pi }{2}(\alpha
-1)\right) Im F(s,0) + \nonumber
\\
+ s^{\alpha - 1}\frac{\pi }{2}\sec
^{2}\left( \frac{\pi }{2}(\alpha -1)\right) \frac{d}{d\ln s}\left( \frac{
Im F(s,0)}{s^{\alpha - 1}}\right).
\end{eqnarray}

At last, from equation (1), we obtain the {\it general} relation connecting
$\rho$ and $\sigma_{tot}$:

\begin{equation}
\rho (s)= \frac{4\pi K}{s\sigma_{tot}(s)} + \tan \left( \frac{\pi }{2}(\alpha
-1)\right) + \frac{\pi }{2}\frac{s^{\alpha - 1}}{\sigma_{tot} (s)}\sec
^{2}\left( \frac{\pi }{2}(\alpha -1)\right) \frac{d}{d\ln s}\left( \frac{
\sigma_{tot} (s)}{s^{\alpha - 1}}\right).
\end{equation}

The standard form referred to in the literature and applied to hadronic
scattering corresponds to the particular choice $\alpha = 1$ (but usually
without the subtraction constant) \cite{matthiae,blockcahn,alfa1,kawasaki},
which we shall refer as the {\it conventional} form of the DAR:
\begin{equation} \rho (s) = \frac{4\pi K}{s\sigma_{tot}(s)} +
\frac{1}{\sigma_{tot} (s)}\frac{\pi}{2}\frac{d}{d\ln s}\left( \sigma_{tot}
(s)\right). \end{equation}

In the next section we use both the {\it conventional} and the {\it general}
relations, in order to determine $\rho (s)$ from different parametrizations for
$\sigma_{tot}^{pp}(s)$ and ensembles of data.

\section{Fits and results}
\vspace{0.3cm}

In this section we first present the fits to the total cross sections
for both ensembles defined in Sec. II and then the predictions for $\rho(s)$
obtained by means of DAR. The discussion on all the obtained results is the
content of Sec. IV.

\vspace{0.3cm}
\subsection{\bf  Fits to total cross section}
\vspace{0.3cm}

For each ensemble defined in Sec. II, we fit the data with some standard and
suitable parametrizations for the total cross sections:

\begin{equation}
fit\ 1 : \qquad \sigma_{tot}= A+B\ln s +C(\ln s)^{2}
\end{equation}

\begin{equation}
fit\ 2 : \qquad \sigma_{tot}= A+B\ln s +C(\ln s)^{D}
\end{equation}

\begin{equation}
fit\ 3 : \qquad \sigma_{tot}= A+B\ln s +C(\ln s)^{2}+Rs^{-1/2}
\end{equation}

\begin{equation}
fit\ 4 : \qquad \sigma_{tot}= A+B\ln s +C(\ln s)^{D}+Rs^{-1/2}
\end{equation}
where $A,\ B,\ C,\ D\ $ and $R$ are free parameters. 

The choice for these parametrizations was based on the following
considerations. Firstly, since Figure 1 is a linear-log plot, we see that 
at the highest energies the
data suggest an increase of \s as a polynomial on $ln s$. The
Froissart-Martin bound states that the fastest rate permissible asymptotically
for the rising of \s is $ln^2 s$ \cite{froissart}. However, it has been shown
by the UA4/2 Collaboration that fits to $pp$ and $\overline{p}p$ accelerator
data indicate the power $2.25^{+ 0.35}_{- 0.31}$, a result referred to as a
``qualitative saturation of the Froissart-Martin bound'' \cite{ua4/2}. For
these reasons, we consider polynomial functions on $ln s$ with two
possibilities for
the power factor: the value $2$, according to the asymptotical bound and as
a free parameter to be determined by the fits. Finally, the power function on
$s$, Eqs. (24) and (25), represents the usual way to take account of data at
lower values of the energy ($5 \sim 20$ GeV) and can be associated with
Regge phenomenology.

The fits were performed by using the CERN-MINUIT routine \cite{minuit} and
the results for ensembles A and B are shown in Figs. 2 and 3, respectively.
The corresponding values of the free parameters and the $\chi^2$ per degree of freedom
are displayed in Table 1 and 2 for ensembles A and B, respectively. We discuss
all these results in Sec. IV.

\begin{minipage}{17.0cm}
\begin{table}
\begin{center}
Table 1. Values of the parameters in Equations $(22)\ -\ (25)$
and the $\chi^2$ per degree of freedom in each fit to Ensemble A.
\vspace{0.5cm}
\begin{tabular}{ccccc}
\hline
fit:     &     1     & 2       &      3     & 4 \\
\hline
$A$      & 45.78     & 42.64   & 38.74      & 33.67  \\
$B$      & -3.315    & -1.875  & -1.868     & -0.3201  \\
$C$      &  0.3654   &  0.1220 &  0.2894    &  0.09465 \\
$D$      & 2 (fixed) &  2.312  &  2 (fixed) &  2.288 \\
$R$      &     -     &    -    & 21.52      & 31.40 \\
$\chi^2$ &   11.7    & 15.4    &  9.8       &  9.9  \\
$d.o.f.$ &   11      & 10      & 10         &   9 \\
$\chi^2/d.o.f.$ & 1.06 & 1.54 & 0.98 & 1.10  \\
\hline
\end{tabular}
\end{center}
\end{table}
\end{minipage}

\begin{minipage}{17.0cm}
\begin{table}
\begin{center}
Table 2. Values of the parameters in Equations $(22)\ -\ (25)$
and the $\chi^2$ per degree of freedom in each fit to Ensemble B.
\vspace{0.5cm}
\begin{tabular}{ccccc}
\hline
fit:     &     1     & 2         &      3     & 4 \\
\hline
$A$      &  49.27    &  43.54    &  62.48     &  38.34 \\
$B$      &  -4.435   &  -1.846   &  -7.221    &  -0.8162\\
$C$      &   0.4534  &   0.05958 &   0.6050   &   0.03461\\
$D$      & 2 (fixed) &  2.624    &  2 (fixed) &   2.755 \\
$R$      &     -     & -         & -38.40     &  14.80\\
$\chi^2$ &  15.0     & 10.7      &  10.4      &  11.4 \\
$d.o.f.$ &  11       &  10       &  10        &   9\\
$\chi^2/d.o.f.$ & 1.36 & 1.07 & 1.04 & 1.27 \\
\hline
\end{tabular}
\end{center}
\end{table}
\end{minipage}

\begin{figure}
\begin{center}
\epsfig{figure=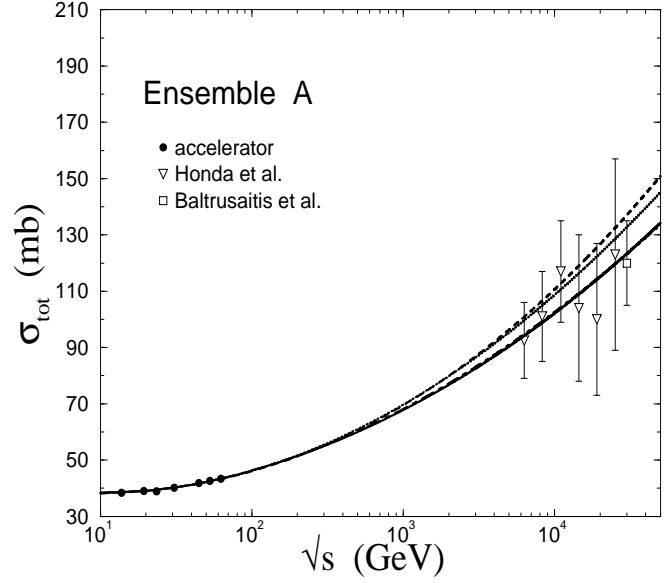,width=10cm,height=9cm}
\vspace{0.5cm}
\caption{Fits to \protect$pp$ total cross section data from ensemble A
through Eqs.
(22) - (25): fit 1 (solid), fit 2 (dotted), fit 3 (dot-dashed ) and fit 4
(dashed).}
 \end{center}
\end{figure}

\begin{figure}
\begin{center}

\epsfig{figure=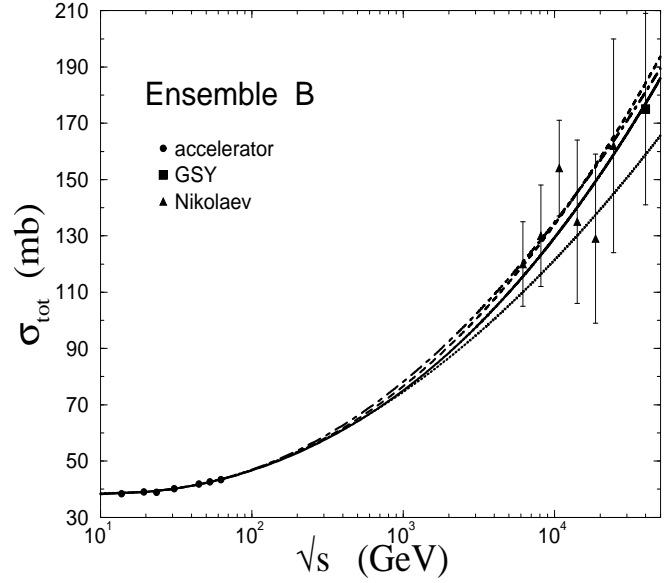,width=10cm,height=9cm}
\vspace{0.5cm}
\caption{Fits to \protect$pp$ cross section from Ensemble B. Same legend as
Fig. 2.} \end{center}
\end{figure}

\vspace{0.3cm}
\subsection{\bf Predictions for the $\rho$ parameter}
\vspace{0.3cm}

In this section we determine the $\rho(s)$ behavior, making use of the DAR
and all the parametrizations for \s obtained in the last section.
As commented before we are interested in the practical applicability of both
forms of the DAR, the  {\it general} and the {\it conventional}
ones, Eq. (20) and Eq. (21), respectively. 

In principle, the {\it general} expression (20) has
two ``free parameters'', namely the subtraction constant $K$ and the parameter
$\alpha$.
Firstly, as a simple exercise to see the effect of the DAR, let us consider
one of the fits to \s, for example fit 4, for both ensembles A and B, and
calculate $\rho(s)$ through the {\it conventional} expression and without
subtraction constant, which means taking $\alpha = 1$ and $K = 0$ in Eq. (20),
or $K = 0$ in Eq. (21) (this corresponds to the formula usually referred to in
the literature \cite{matthiae,blockcahn}).
The results are displayed in Fig. 4 together with the experimental data
\cite{rho}. We see that although the predictions from both ensembles are
similar at the ISR energy region, both disagree with the data.

In what follows we investigate the influence of the above two free parameters
in the description of these data.
In order to do that, we first treat the {\it
conventional} formula with the subtraction constant and then the {\it general}
formula, without subtraction constant and  the factor $\alpha$ as free
parameter.

\vspace{0.3cm}
$\bullet$ {\it Conventional} derivative dispersion relation
\vspace{0.3cm}

With the parametrizations (22-25) for \s we obtain analytical
expressions for $\rho (s)$ by using the {\it conventional} form
of the DAR, Eq. (21). For each input parametrization the subtraction
constant $K$ is a free parameter, determined by fit, through the CERN-MINUIT
routine, to the experimental $\rho$ data \cite{rho}.  

The results from each parametrization and for both ensembles are shown
in Figures 5 and 6 together with the experimental data. The values of the
subtraction constant and statistical information about the fits are displayed
in Table 3.

\begin{figure}
\begin{center}

\epsfig{figure=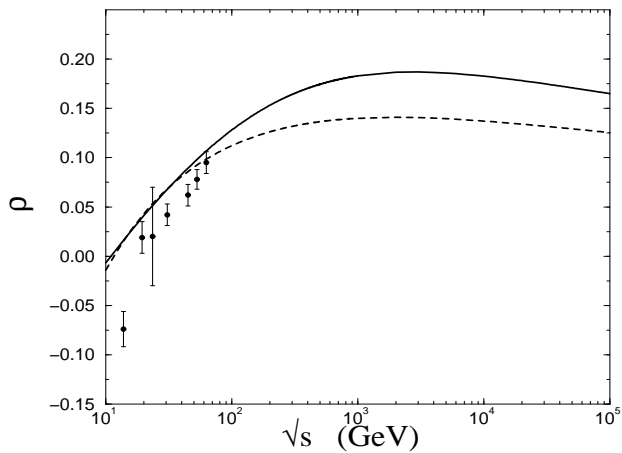,width=9cm,height=7cm}
\vspace{0.5cm}
\caption{ Results for \protect$\rho(s)$, using fit 4 from both ensembles
A (dashed) and B (solid) and DAR with \protect$\alpha = 1$ and \protect$K = 0$
in Eq. (20) and experimental data \protect\cite{rho}.}
 \end{center}
\end{figure}

\begin{minipage}{17.0cm}
\begin{table}
\begin{center}
Table 3. Values of the subtraction constant from fits to \protect$\rho(s)$
data and the $\chi^2$ for 6 degree of freedom. Calculation performed thorough
Eq. (21)  using the four fits to \s from ensembles A and B.
\vspace{0.3cm}
\begin{tabular}{ccccc}
\hline
fit-\s- Ensemble A:             &     1     & 2         &      3     & 4    \\
\hline
$ K $ &  -134.1   &  -135.6   &  
-130.3   &  -129.6 \\ $\chi^2/d.o.f.$     &    0.82   &    0.63   &   0.97   &
0.63 \\ \hline
fit-\s- Ensemble B:             &     1     & 2         &      3     & 4    \\
\hline
$ K $ &   -130.0 & -134.4   &  -137.1   & -132.0  \\
$\chi^2/d.o.f.$     &    2.0 &    1.48  &   1.85  & 1.44\\
\hline
\hline
\end{tabular}
\end{center}
\end{table}
\end{minipage}

\begin{figure}
\begin{center}

\epsfig{figure=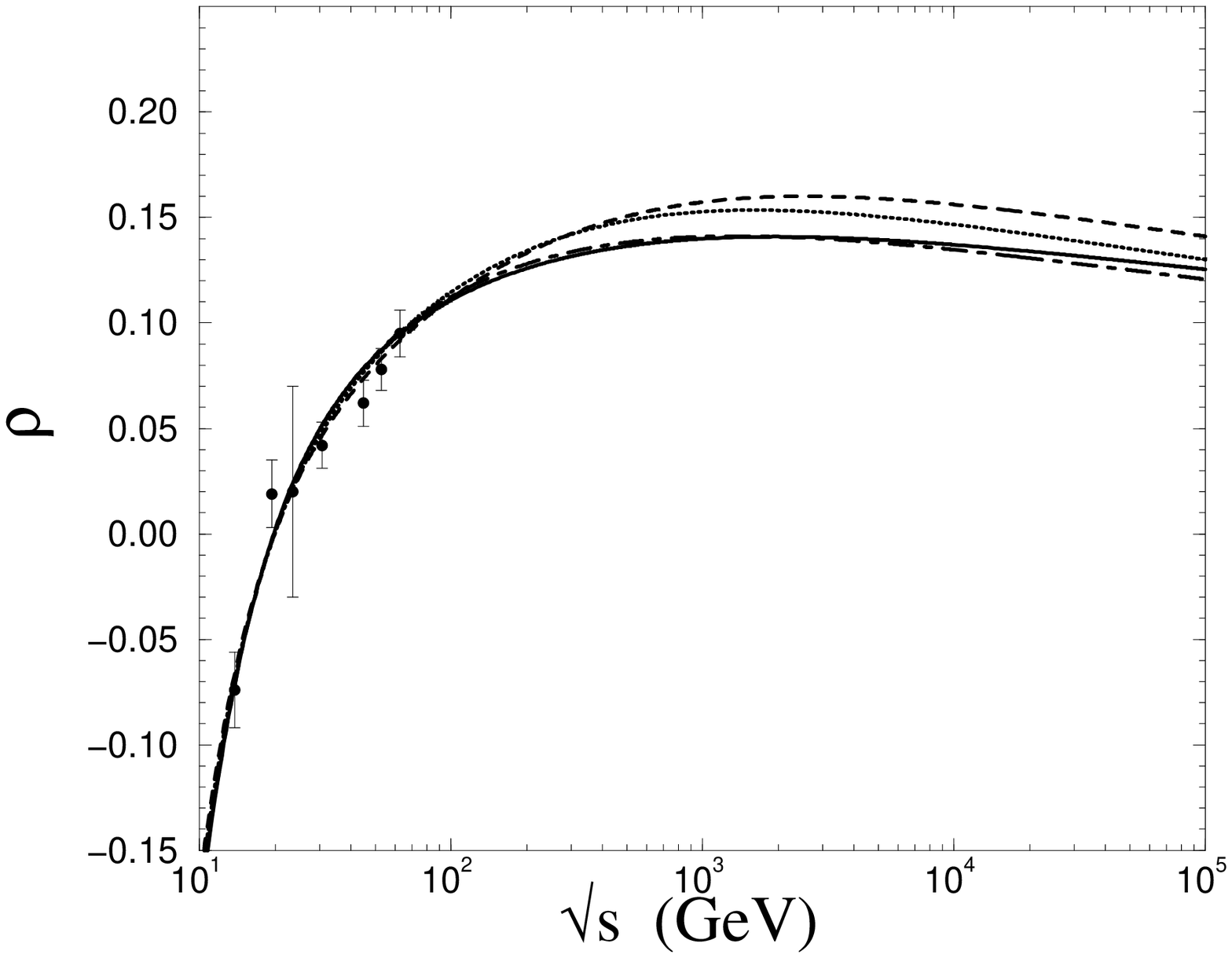,width=10cm,height=8cm}
\caption{Results for $\rho (s)$, through the {\it conventional} DAR, Eq.
(21),  from fits to \protect\s from ensemble A together with the
experimental data. Same legend as Fig. 2.}
\end{center}
\end{figure}

\begin{figure}
\begin{center}

\epsfig{figure=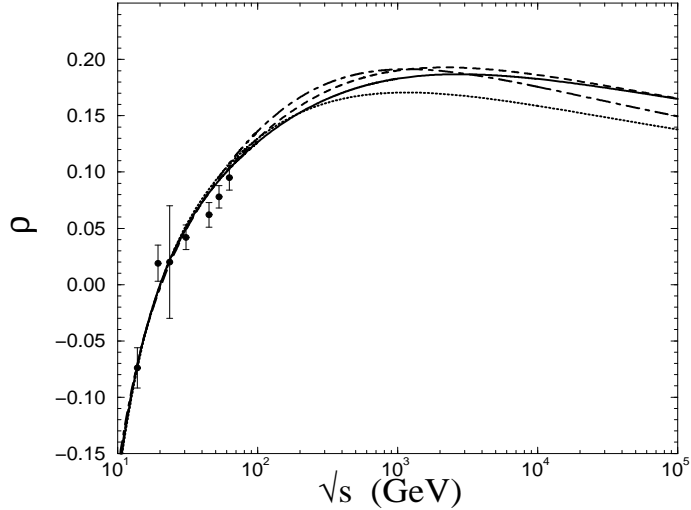,width=10cm,height=8cm}
\caption{Results for $\rho (s)$, through the {\it conventional} DAR, Eq. (16), 
fits to \protect\s from ensemble B, together with the experimental
data. Same legend as Fig. 2. }
\end{center}
\end{figure}

\vspace{0.3cm}
$\bullet$ {\it General} derivative analyticity relation
\vspace{0.3cm}

As shown in some detail in Sec. II.B, the general result for the DAR depends on
the free parameter $\alpha$.  It comes from the integration by parts of Eq.
(7) and it is necessary in order to allow a finite derivative form associated
with the integral form. In this sense it seems to play the role of a
regularization factor. Although in nearly all practical uses of the DAR the
value $\alpha = 1$ is assumed, we have shown that its value is constrained to
the interval $0 < \alpha <2$ (see also references
\cite{crit-dar,kolarfischer}). In particular, in the context of a multiple
diffraction model, it was recently shown that  the description of the
experimental data on $pp$ elastic scattering may be improved by taking
$\alpha$ as a free parameter \cite{mmpsn}. This early result inspired us to
make use of the general relation (20) and to investigate the possible effect
of $\alpha$ as a free parameter (some preliminary results were already
presented in Ref. \cite{saolourenco00}). To this end we will not take account
of the subtraction constant in the general formula for the DAR, namely $K = 0$
in Eq. (20), so that we explicitly have:

\begin{eqnarray}
\rho (s)&=&\tan \left( \frac{\pi }{2}(\alpha
-1)\right) + \nonumber
\\
&+& \frac{\pi }{2}\sec
^{2}\left( \frac{\pi }{2}(\alpha -1)\right)\left[ \frac{1}{\sigma_{tot}(s)} 
\frac{d\sigma_{tot}(s)}{d\ln s} + 1 - \alpha \right].
\end{eqnarray}

With a fixed parametrization for \s we can fit the experimental data
on $\rho$ by letting $\alpha$ to be a free parameter in the above
equation, once more by using the CERN-MINUIT.  The results from all the
parametrizations for \s with both ensembles A and B are shown in Figures 7 an
8, respectively. The corresponding values for $\alpha$ and statistical
information about the fits are displayed in Table 4.

\begin{minipage}{17.0cm}
\begin{table}
\begin{center}
Table 4. Values of the parameter \protect$\alpha$ from fits to
\protect$\rho(s)$ data and the $\chi^2$ for 6 degree of freedom. Calculation
performed thorough Eq. (20) with \protect$K = 0$ and  using the four fits to
\s from ensembles A and B. \vspace{0.5cm}
\begin{tabular}{ccccc}
\hline
fit-\s- Ensemble A: &     1     & 2         &      3     & 4    \\
\hline
$\alpha$         &   1.229   & 1.223   &  1.231   & 1.230  \\
$\chi^2/d.o.f.$  &   2.61   &  3.17    &  2.17   &  2.20\\
\hline
fit-\s- Ensemble B: &     1     & 2         &      3     & 4    \\
\hline
$\alpha$      &  1.244 & 1.240  & 1.244   & 1.239 \\
$\chi^2/d.o.f.$ &  1.48  & 1.94    &  2.00  & 1.80 \\
\hline
\end{tabular}
\end{center}
\end{table}
\end{minipage}

\begin{figure}
\begin{center}

\epsfig{figure=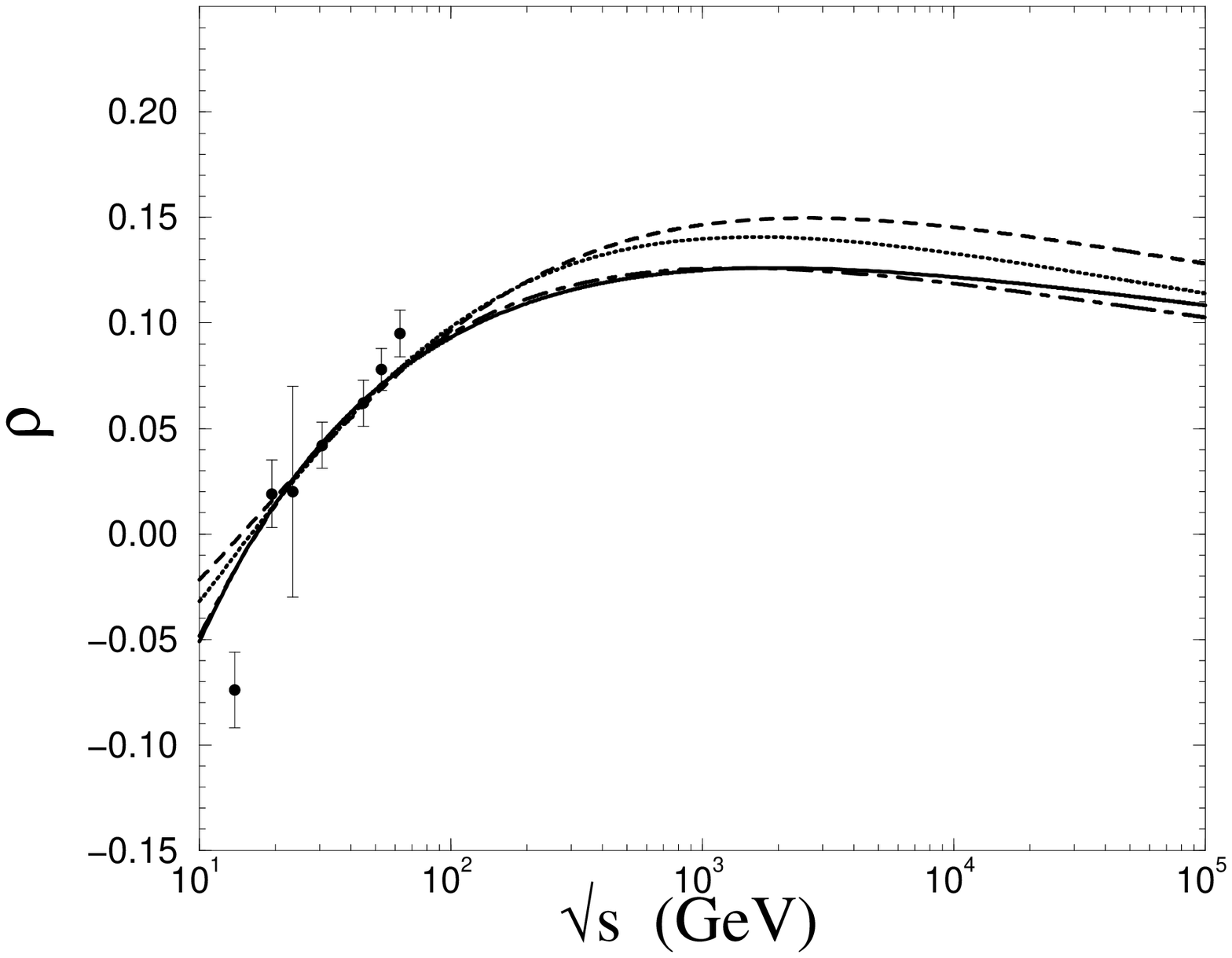,width=10cm,height=9cm}
\vspace{0.5cm}
\caption{ Results for \protect$\rho (s)$, through the general expression of
the  DAR, Eq. (20), with \protect$\alpha$ as free parameter, \protect$K = 0$
and  fits to \protect\s from ensemble A. } 
\end{center}
\end{figure}

\begin{figure}
\begin{center}

\epsfig{figure=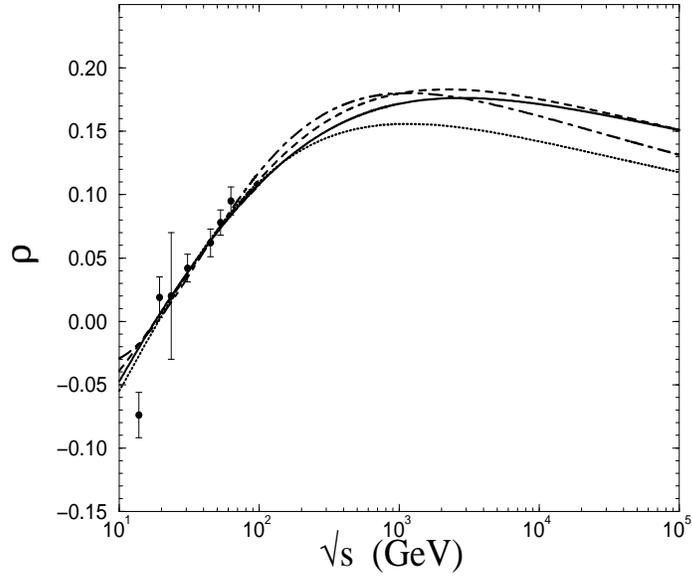,width=10cm,height=9cm}
\vspace{0.5cm}
\caption{Results for \protect$\rho (s)$, through the general expression of the 
DAR, Eq. (20), with \protect$\alpha$ as free parameter, \protect$K = 0$ and 
fits  to \protect\s from ensemble B.}  
 \end{center}
\end{figure}


\section{Discussion}

The parametrizations for \s are  the usual ones, but the different ensembles,
suggested by cosmic-ray results, introduce novel behaviors in the asymptotic
region, as clearly shown in Figs. 2 and 3. For example, by calculating the
average value and the standard deviation from the four parametrizations we can
estimate $\sigma_{tot}^{pp}(14\ \textnormal{TeV}) = 113 \pm 5$ mb for ensemble
A and  $\sigma_{tot}^{pp}(14\ \textnormal{TeV}) = 140 \pm 7$ mb for ensemble
B. At lower values of the energy, namely $\sqrt s$ in the region $10\ - \ 100$
GeV, there is no significant distinction between the four parametrizations.
In general, model predictions are in agreement with the above result from
ensemble A \cite{dl,dgp,bsw}, including the fit by the UA4/2 Collaboration
\cite{ua4/2}. To our knowledge, the only exception (model) that presents
agreement with the result from ensemble B is that of Ref. \cite{mm}.
This multiple diffraction model is based on analyzes of $pp$ elastic scattering
in the interval $13.8\, \textnormal{GeV} \le \sqrt s \le 62.5$ GeV, the
same set we used here at the accelerator region. Extrapolation to higher
energies predicted $\sigma_{tot}(16\ \textnormal{TeV}) = 147$ mb, without
estimated errors.  Recently P\'erez-Peraza et al. improved this model
predictions determining confident error bands through a forecasting regression
analysis \cite{velasco}. Reading from Fig. 2 of this reference we can infer 
$\sigma_{tot}(16\ \textnormal{TeV}) \sim 147 \pm 37$ mb. Despite of the large
error band ( $\sim 25\% $), even in this case it is clear from the quoted
figure that the results favors ensemble B.
From a ``statistical point of view'' we may say that the published results
from models and fits show agreement with ensemble A. However, it should be
stressed that the cosmic-ray estimations in the ensemble A were obtained under
the hypothesis of the geometrical scaling \cite{engel}, which is violated even
at $ \sqrt s \sim 500$ GeV.
Moreover, we
should remember results from cosmic-ray experiments which indicate
the possibility of new phenomena in $pp$ collisions at center-of-mass energies
beyond $500$ GeV \cite{centauro}. As a direct consequence we should expect
new open channels and therefore a faster rising of the $pp$ total cross
section than expected in the extrapolations from accelerator data. This seems
to be well accommodated by the parametrizations with ensemble B.

Concerning the determination of $\rho(s)$ through DAR, we first showed that
the {\it conventional} expression without subtraction constant ($\alpha = 1$
and  $K = 0$) does not reproduce the experimental $\rho$ data (Fig. 4). We
recall that this expression has been referred to in the literature
\cite{matthiae,blockcahn} and also used in the context of phenomenological
models \cite{alfa1,kawasaki}. Taking account of the subtraction constant $K$,
the data are well described, as shown in Figs. 5 and 6.
From Table 3, 
\begin{equation}
 |K| \sim 130 - 137,
\end{equation}
depending on the fit and ensemble for \s. As expected different ensembles
correspond to distinct behaviors at the highest and asymptotic energies. As in
the case of \s we can estimate an average value and standard deviation from the
four results in Figs. 5 and 6: $\rho(14\ \textnormal{TeV}) = 0.142 \pm 0.010$
with \s from ensemble A and $\rho(14\ \textnormal{TeV}) = 0.173 \pm 0.013$ from
ensemble B. In this case we can say that the results with ensemble A present
the best agreement with the experimental data (Table 3 and Figs. 5 and 6).
We also tested the {\it general} expression for the DAR by taking $K = 0$
and letting $\alpha$ as a free fit parameter. The results presented in
Figs. 7 and 8 show that the data are also satisfactorily described, specially
in the case of \s from ensemble B (compare with Fig. 4, the case of fixed
$\alpha = 1$). From Table 4,

\begin{equation}
\alpha \sim 1.22 - 1.24,
\end{equation}
depending also on the fit and ensemble for \s. In all these cases the
condition (13) is verified.

\section{Conclusions and final remarks}
In this communication we investigated two ensembles of experimental
information on $pp$ total cross sections and used four different
parametrizations to fit the data, as function of the energy.
In each case we obtained predictions for $\rho (s)$ making use of both the
{\it conventional} and the {\it general} expressions for the derivative
analyticity relations.

Our first main conclusion is that experimental information presently available
on $pp$ total cross sections indicates two possible different scenarios for
the hadronic interactions at the highest energies. The fast rise of \s
from the analysis of ensemble B is corroborated by the multiple
diffraction model of Ref. \cite{mm} and  also indication of new phenomena
from emulsion chamber experiments \cite{centauro}.
Although in this work we only point out for this possibility, new information
coming from RHIC and LHC shall certaintly clarify this subject. In this sense,
our results may be viewed as a kind of warning against some possible
precipitated assumptions, namely,  that extrapolations from accelerator data
which show agreement with the cosmic-ray estimations in ensemble A could be
the final answer to the question.

A second novel result from this model independent analysis 
was to show the practical applicability of both the {\it
conventional} and the {\it general} expressions for the derivative analyticity
relations at sufficiently high energies. In the {\it conventional} case,
$\alpha = 1$, letting the constant $K$ as free parameter, the description of
the experimental $\rho$ data is quite good. On the other hand, taking $K = 0$
and letting $\alpha$ as free fit parameter, the results may be considered 
satisfactory. In both cases the predictions at the highest energies are
practically the same (compare Figs. 5 and 7 and also Figs. 6 and and 8) and the
differences at this energy region come obviously from the different ensembles
for \s. Our approach was to consider the two possibilities separately ($K \not=
0$ or $\alpha \not= 1$), so that we could infer the intervals of possible
variations, Eqs. (27) and (28). Simultaneous analysis with both possibilities
shall improve the description of $\rho(s)$ and this is our second main
conclusion.

In this communication we treated only $pp$ interactions, since the
cosmic-ray informations concern only this case. We are presently
investigating the inclusion of antiproton-proton data in the analysis
through adequate considerations on crossing symmetry.


We are grateful to Fapesp for financial support and J. Montanha for reading the
manuscript. M.J.M. is also  thankful  to N.N. Nikolaev for 
discussions, V. Ezhela for correspondence and CNPq for financial
support.

\end{document}